\documentclass[stslayout]{imsart}
\usepackage{geometry} 
\usepackage{graphicx}	
\usepackage{amssymb, amsmath}
\usepackage{epstopdf}
\usepackage{natbib}
\usepackage{subfigure}
\usepackage[colorlinks,citecolor=blue,urlcolor=blue]{hyperref}

\DeclareFontFamily{U}{mathx}{\hyphenchar\font45}
\DeclareFontShape{U}{mathx}{m}{n}{
      <5> <6> <7> <8> <9> <10>
      <10.95> <12> <14.4> <17.28> <20.74> <24.88>
      mathx10
      }{}
\DeclareSymbolFont{mathx}{U}{mathx}{m}{n}
\DeclareFontSubstitution{U}{mathx}{m}{n}
\DeclareMathAccent{\widecheck}{0}{mathx}{"71}

\newcommand\dd{\mathrm{d}}

\newcommand{\sLS}{\ensuremath { H^{-1} \left( E \right) } }

\graphicspath{{figures/}}

\begin{document}

\begin{frontmatter}

\title{Diagnosing Suboptimal \\
        Cotangent Disintegrations \\
        in Hamiltonian Monte Carlo}
\runtitle{Diagnosing Suboptimal Cotangent Disintegrations}

\begin{aug}
  \author{Michael Betancourt%
  \ead[label=e1]{betanalpha@gmail.com}}

  \runauthor{Betancourt}

  \address{Department of Statistics, University of Warwick, 
  Coventry CV4 7AL, UK \\ \printead{e1}.}

\end{aug}

\begin{abstract}
When properly tuned, Hamiltonian Monte Carlo scales
to some of the most challenging high-dimensional problems
at the frontiers of applied statistics, but when that tuning is
suboptimal the performance leaves much to be desired.
In this paper I show how suboptimal choices of one critical
degree of freedom, the cotangent disintegration, manifest 
in readily observed diagnostics that facilitate the robust
application of the algorithm.
\end{abstract}

\end{frontmatter}

Once a statistical model has been specified as a probability distribution, 
applied statistics reduces to the evaluation of expectations with respect to the 
that target distribution.  Consequently, the fundamental computational challenge 
in these statistics is the accurate and efficient estimation of these expectations.  

Given its general applicability, Markov chain Monte Carlo~\citep{RobertEtAl:1999,
BrooksEtAl:2011} has become one of the most of the most popular frameworks 
for developing practical estimation algorithms, as evident from decades of 
theoretical analysis and empirical success.  In particular, Hamiltonian Monte 
Carlo~\citep{DuaneEtAl:1987, Neal:2011, BetancourtEtAl:2014a} pushes Markov 
chain Monte Carlo deep into the frontiers of applied statistics by exploiting the
geometry inherent to many probability distributions.  Implementing Hamiltonian 
Monte Carlo in practice, however, is frustrated by algorithmic degrees of freedom
that present a delicate tuning problem which can not only impede the scalable
performance of the algorithm but also introduce biases in the estimation.  

In this paper I consider the choice of a \emph{cotangent disintegration} that 
arises in any Hamiltonian Monte Carlo algorithm.  Because the performance of 
the resulting implementation is highly sensitive to the interaction of the cotangent
disintegration with the given target distribution, a careful choice is critical for 
robust performance.

After first reviewing the general construction of Hamiltonian Monte Carlo, I
show how the consequences of a given cotangent disintegration manifest
in the performance of a single stage of the algorithm.  I then analyze this 
stage to define not only a implicit criteria for the optimal disintegration 
relative to a given target distribution, but also an explicit diagnostics to 
identify a suboptimal cotangent disintegration in practice.  Finally I
demonstrate the utility of these diagnostics in various examples.

\section{Constructing Hamiltonian Monte Carlo}

In this paper I will let $\pi$ be the target probability distribution over the 
$D$-dimensional sample space $Q$ and some appropriate $\sigma$-algebra.  
To simplify the notation I will assume that $\pi$ admits a density 
$\pi \! \left( q \right)$ with respect to some reference measure, $\dd q$, 
although Hamiltonian Monte Carlo does not require this.  For the more 
general construction of Hamiltonian Monte Carlo see 
\cite{BetancourtEtAl:2014a}.

Here I will very briefly review Markov chain Monte Carlo and then
Hamiltonian Monte Carlo, both in general and in its most common
implementation.

\subsection{Markov Chain Monte Carlo}

Markov chain Monte Carlo builds expectation estimates by finding and
exploring the neighborhoods on which the target probability distribution
concentrates.  The exploration itself is generated by repeatedly sampling 
from a Markov transition, given by the density 
$\mathcal{T} \! \left( q \mid q' \right)$, to give a sequence of points, 
$\{ q_{0}, \ldots, q_{N} \}$, known as a Markov chain.  If the transition 
preserves the target distribution,
\begin{equation*}
\pi \! \left( q \right) = 
\int_{Q} \mathcal{T} \! \left( q \mid q' \right) \pi \! \left( q' \right) \dd q',
\end{equation*}
then the resulting Markov chain will eventually explore the entire target 
distribution and we can use the history of the Markov chain to construct
consistent Markov chain Monte Carlo estimators of the desired
expectations,
\begin{equation*}
\lim_{N \rightarrow \infty} \hat{f}_{N}
\equiv
\lim_{N \rightarrow \infty} \frac{1}{N} \sum_{n = 0}^{N} f \! \left( q_{n} \right)
= \mathbb{E}_{\pi} \! \left[ f \right].
\end{equation*}

The performance of these Markov chain Monte Carlo estimators depends
on how effectively the Markov transition guides the Markov chain 
along the neighborhoods of high probability.  If the exploration is
slow then the estimators will become computationally inefficient,
and if the exploration is incomplete then the estimators will become
biased.  In order to scale Markov chain Monte Carlo to the
high-dimensional and complex distributions of practical interest,
we need a Markov transition that exploits the properties of the
target distribution to make informed jumps through neighborhoods
of high probability while avoiding neighborhoods of low probability
entirely.

\subsection{Hamiltonian Monte Carlo}

Hamiltonian Monte Carlo achieves such informed transitions by
harnessing the differential geometry of the target distribution with
auxiliary \emph{momenta} parameters.  The algorithm begins by
first attaching to each point, in the sample space, $q$, a copy of
$\mathbb{R}^{D}$ called a \emph{momenta fiber}.  Collecting these 
fibers together yields the $2D$-dimensional \emph{cotangent bundle}, 
$T^{*} Q$, with a natural projection that collapses each fiber to the 
base point at which it was attached, 
\begin{align*}
\varpi:& \; T^{*} Q \rightarrow Q
\\
& (q, p) \mapsto q.
\end{align*}

We next lift our target probability distribution into a joint distribution
on the cotangent bundle with the choice of a conditional probability 
distribution over the fibers known as a \emph{cotangent disintegration}.  
Denoting the target distribution 
\begin{equation*}
\pi \propto \exp \! \left( - V ( q ) \right) \dd q,
\end{equation*}
with $V ( q )$ denoted the \emph{potential energy}, and the cotangent 
disintegration as
\begin{equation*}
\xi_{q} \propto \exp \! \left[ - K (q, p) \right] \dd p,
\end{equation*}
with $K (q, p)$ denoted the \emph{kinetic energy}, then the joint distribution
is defined as
\begin{align*}
\pi_{H} 
&=
\xi_{q} \cdot \pi
\\
&\propto \exp \! \left( - \left( K (q, p) + V (q) \right) \right) \dd q \, \dd p
\\
&\propto \exp \! \left(- H (q, p) \right) \dd q \, \dd p,
\end{align*}
with $H (q, p)$ denoted the \textit{Hamiltonian}.

When combined with the natural fiber structure of the cotangent bundle,
this Hamiltonian immediately defines an infinite family of deterministic 
maps,
\begin{align*}
\phi^{H}_{t} : (q, p) &\rightarrow (q, p), \forall t \in \mathbb{R}
\\
\phi^{H}_{t} \circ \phi^{H}_{s} &= \phi^{H}_{s + t},
\end{align*}
called a \emph{Hamiltonian flow}.  By construction, the Hamiltonian
flow traces through the neighborhoods where the joint distribution
concentrations, while its projection, $\varpi \circ \phi_{t}^{H}$, 
traces through the neighborhoods where the target distribution
concentrates, exactly as desired.

\begin{figure}
\centering
\includegraphics[width=4in]{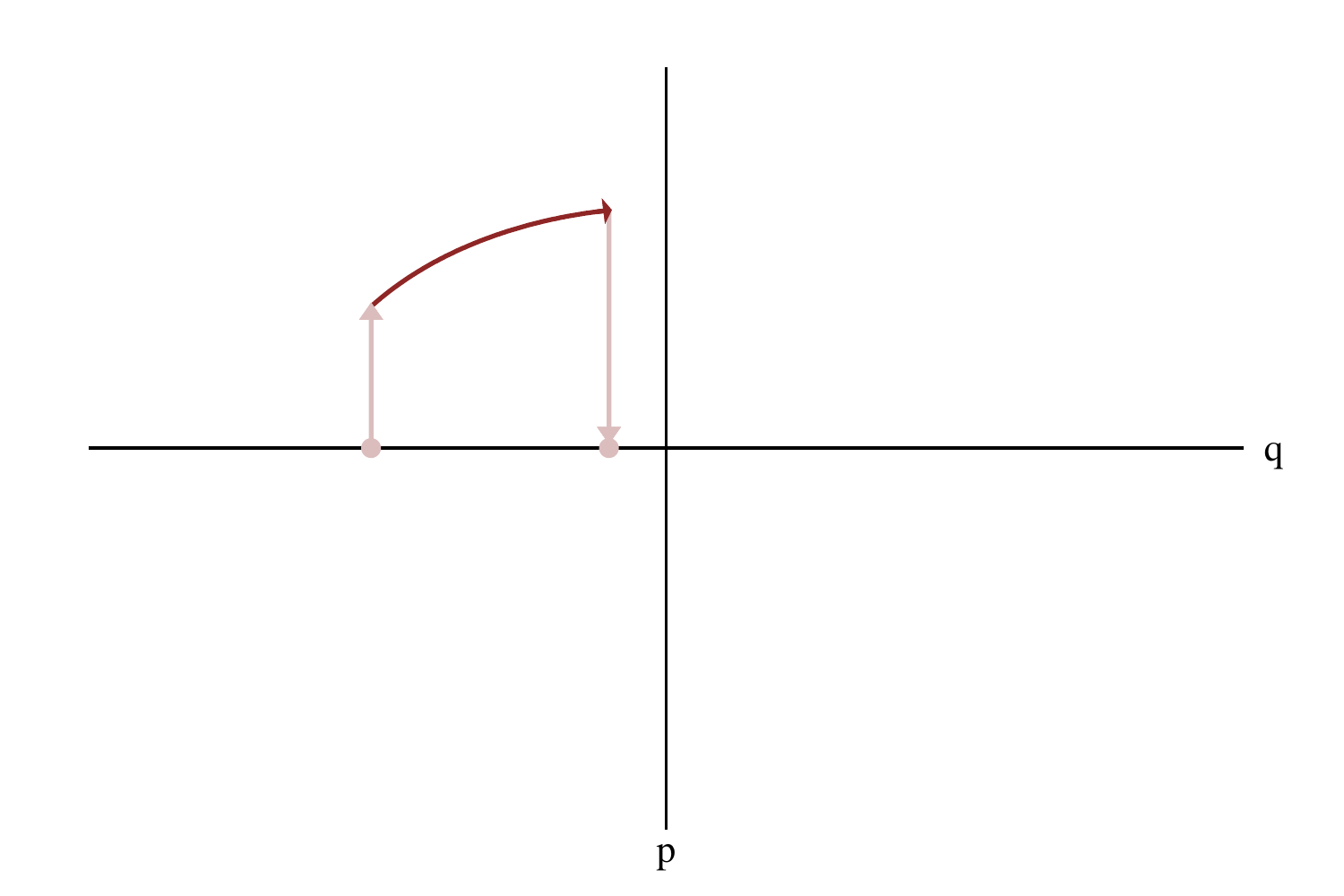}
\caption{Every Hamiltonian Markov transition is comprised of a random
lift from the target sample space onto the cotangent bundle (light red), a 
deterministic Hamiltonian flow on the cotangent bundle (dark red), 
and a projection back down to the target space (light red).}
\label{fig:hmc_transition_cartoon}
\end{figure}

Hence we can build a powerful Markov transition in three stages.
From the initial point, $q$, we first lift from the sample space onto 
the cotangent bundle by sampling a random momenta from the 
cotangent disintegration, $p \sim \xi_{q}$, apply the Hamiltonian flow 
for some time to generate exploration, $(q, p) \mapsto \phi^{H}_{t} (q, p)$, 
and then project back down to the target sample space, 
$\varpi : (q, p) \mapsto q$.  In practice there are various strategies for 
choosing the integration time, as well as numercially approximating 
the Hamiltonian flow and correcting for the resulting 
error~\citep{Betancourt:2016}, but in general any Hamiltonian Markov 
transition will proceed with a lift, a flow, and a projection (Figure
\ref{fig:hmc_transition_cartoon}).

\subsection{Gaussian-Euclidean Cotangent Disintegrations}

An explicit choice of cotangent disintegration is facilitated when the sample
space has a metric structure.  In particular, if the sample space is
equipped with a Riemannian metric, $g$, then we can define an entire 
family of \emph{Riemannian cotangent disintegrations} with the kinetic 
energies
\begin{equation*}
K \! \left( q, p \right) = A \cdot f \! \left( g^{-1}_{q} \! \left( p, p \right) \right)
+ \frac{1}{2} \log \left| g_{q} \right| + \mathrm{const},
\end{equation*}
for some constant $A$ and function $f: \mathbb{R} \rightarrow \mathbb{R}$.  
Riemannian disintegrations also define two helpful scalar functions: 
the \textit{effective potential energy},
\begin{equation*}
\widecheck{V} \! \left( q \right) 
= 
V \! \left( q \right) +  \frac{1}{2} \log \left| g_{q} \right| + \mathrm{const}.
\end{equation*}
and the \textit{effective kinetic energy},
\begin{equation*}
\widecheck{K} \! \left(q, p \right) 
= 
A \cdot f \! \left( g^{-1}_{q} \! \left( p, p \right) \right).
\end{equation*}

In practice most implementations of Hamiltonian Monte Carlo assume
that any metric structure is \emph{Euclidean}, where the metric $g$ is
constant across the sample space and then sometimes denoted as 
a \emph{mass matrix}.  Additionally, these implementations usually take
$A = \frac{1}{2}$ and $f = \mathbb{I}$, in which case the cotangent 
disintegration defines a Gaussian distribution over each momenta fiber.  
Hence in common practice we typically consider only 
\emph{Gaussian-Euclidean cotangent disintegrations}.

\section{The Microcanonical Disintegration}

Although any choice of cotangent disintegration will yield a Hamiltonian 
flow that coherently explores the neighborhoods where the target
distribution concentrates, not every choice will yield a flow that
is as computationally efficient as others.  How the interaction between
a particular disintegration and the target distribution manifests
in performance may at first seem abstruse, but it becomes 
straightforward to characterize if we examining these Hamiltonian 
systems from a more natural perspective.

\begin{figure}
\centering
\subfigure[] {\includegraphics[width=2.9in]{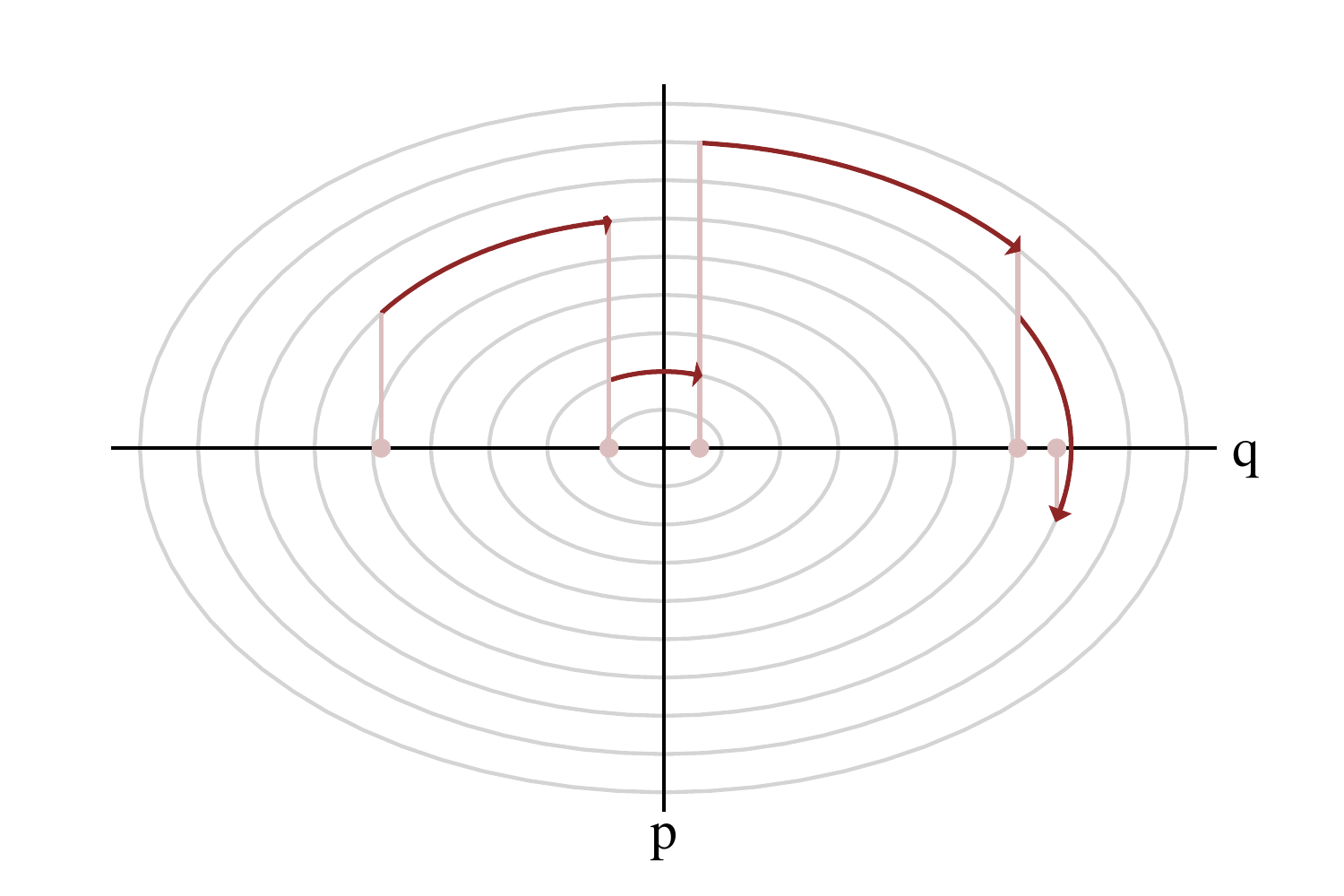} }
\subfigure[] {\includegraphics[width=2.9in]{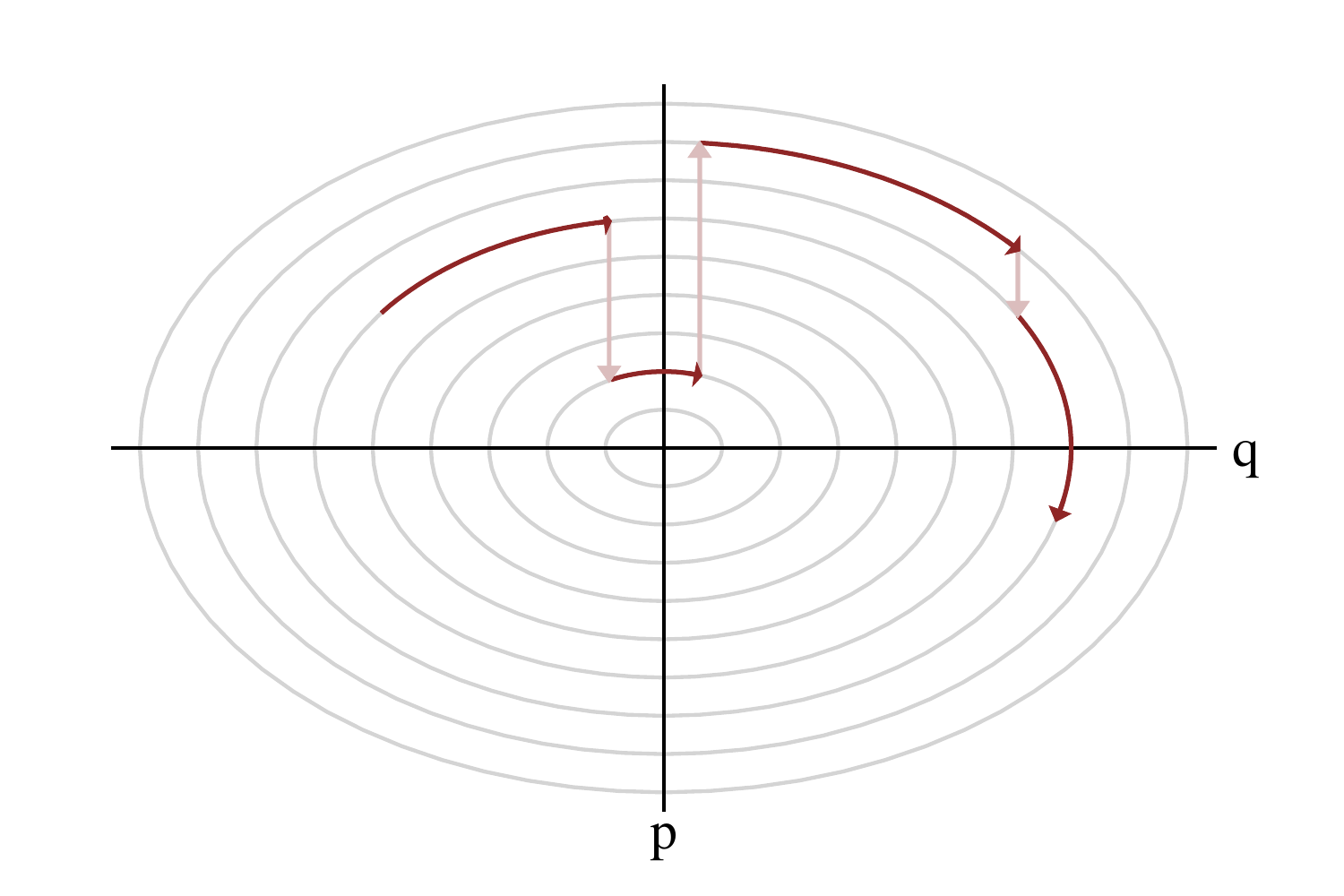} }
\caption{(a) Each Hamiltonian Markov transition lifts the initial
state onto a random level set of the Hamiltonian, which can then
be explored with the Hamiltonian flow before projecting back down
to the target sample space.  (b) If we consider the projection 
and random lift steps as a single momentum resampling operation, 
then the Hamiltonian Markov chain alternates between deterministic 
flows along these level sets (dark red) and a random walk across the 
level sets (light red).}
\label{fig:hmc_chain_cartoon}
\end{figure}

One of the distinctive properties of Hamiltonian flow is that it
preserves the Hamiltonian itself, which implies that each Hamiltonian
trajectory is confined to a \emph{level set} of the Hamiltonian,
\begin{equation*}
H^{-1} \! \left( E \right) 
= 
\left\{ (q, p) \in T^{*} Q \mid H \! \left( q, p \right) = E \right\}.
\end{equation*}
A Markov transition, then, first jumps to a random level set and
then explores that level set with the Hamiltonian flow before
projecting back to the sample space (Figure \ref{fig:hmc_chain_cartoon}a).
If we compose the projection and random lift stages together into a single
\emph{momentum resampling} operation, then the entire Hamiltonian 
Markov chain naturally decouples into exploration along each level set 
driven by the Hamiltonian flow, and exploration across level sets driven
by the momentum resampling (Figure \ref{fig:hmc_chain_cartoon}b).

Consequently a much more natural way to analyze Hamiltonian
Monte Carlo is not through positions and momenta but rather
level sets and the \emph{energies} labeling each level set.  The
\emph{microcanonical disintegration} formalizes this intuition
by decomposing the joint distribution into a conditional 
\emph{microcanonical distribution} over level sets, $\pi_{\sLS}$, 
and a \emph{marginal energy distribution}, $\pi_{E}$,
\begin{equation*}
\pi_{H} = \pi_{\sLS} \cdot \pi_{E}.
\end{equation*}
A Hamiltonian system always admits a microcanonical disintegration,
although there are some technical subtleties~\citep{BetancourtEtAl:2014a}.

From this perspective, the Hamiltonian flow generates exploration 
of the microcanonical distributions while the exploration of the marginal 
energy distribution is determined solely by the momentum resampling.
Because the cotangent disintegration affects the geometry of the level sets,
it also effects the efficacy of the Hamiltonian flow, but this can largely be
overcome with an appropriate choice of integration 
times~\citep{Betancourt:2016}.  The exploration of the marginal energy 
distribution, however, is determined solely by the momentum resampling 
which itself depends on only the interaction between the cotangent 
disintegration and the target distribution.

\section{Diagnosing Suboptimal Cotangent Disintegrations}

\begin{figure}
\centering
\subfigure[]{ \includegraphics[width=2.5in]{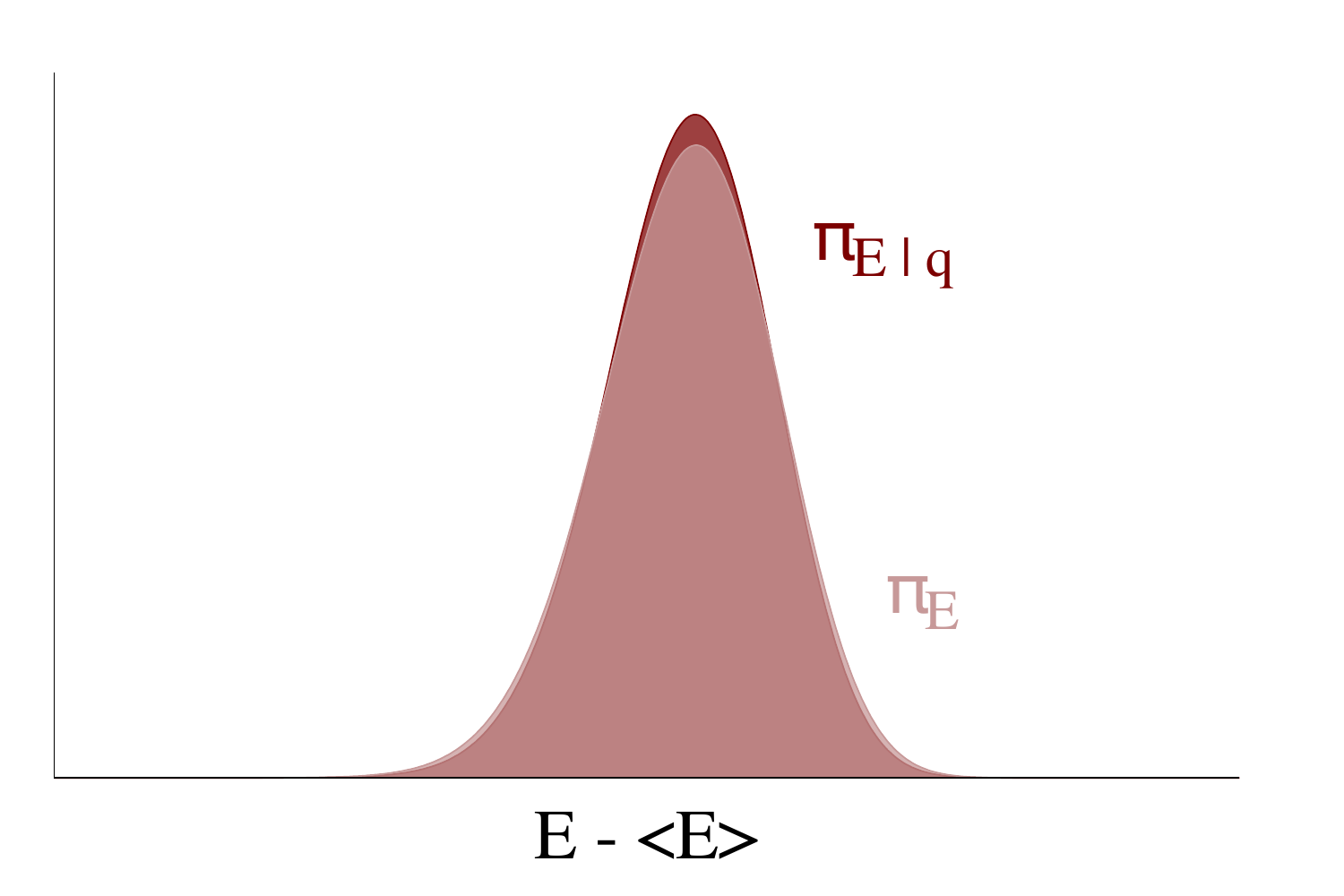} }
\subfigure[]{ \includegraphics[width=2.5in]{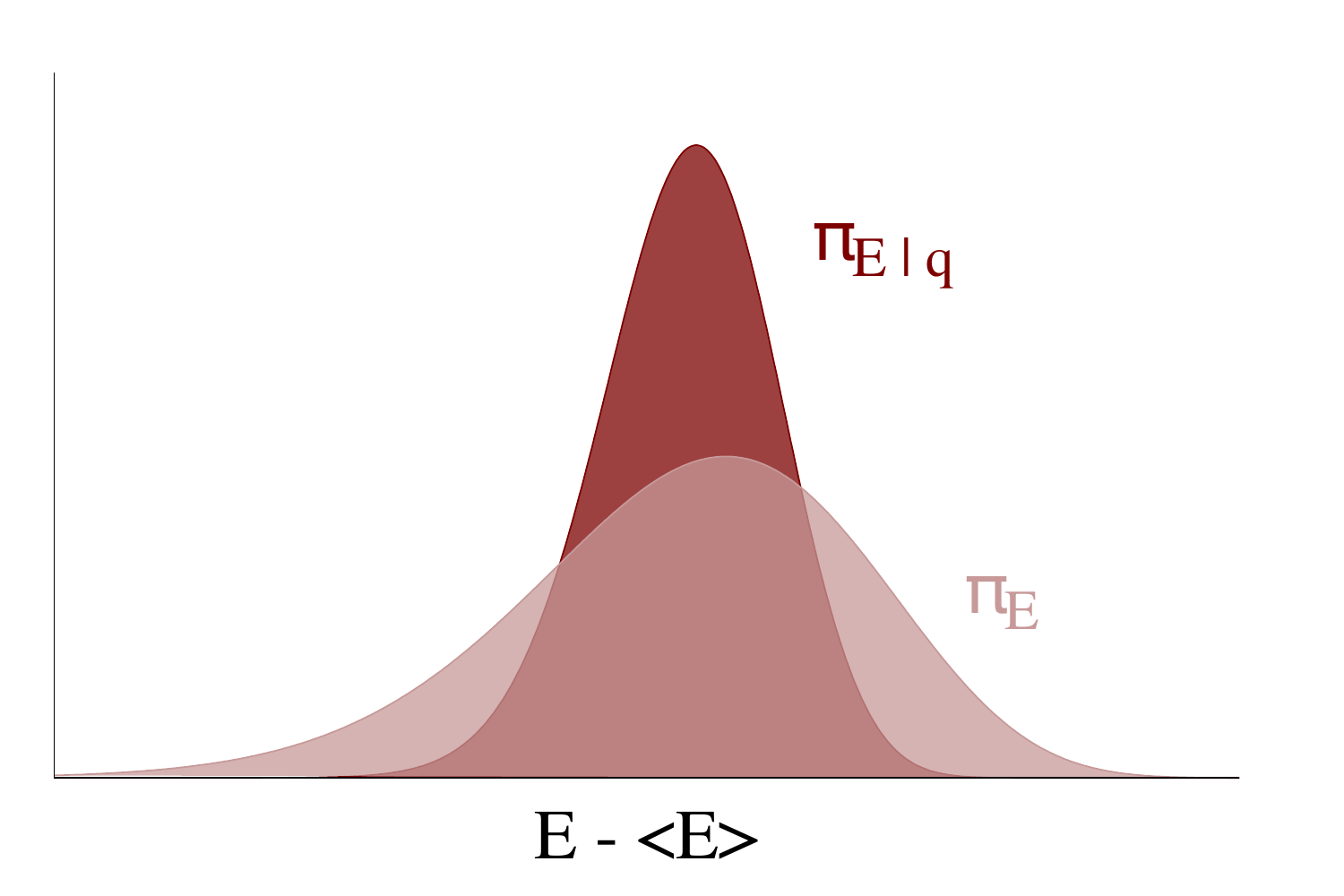} }
\caption{The momentum resampling in a Hamiltonian Markov transition
induces a change of energies and allows a Hamiltonian Markov chain to 
randomly walk across level sets.  (a) When the energy distribution induced
by momentum resampling at any point $q$, $\pi_{E \mid q}$ is similar to the 
marginal energy distribution, $\pi_{E}$, this random walk will rapidly explore 
all relevant energies and the resulting Hamiltonian Markov chain will enjoy
small autocorrelations.  (b) On the other hand, when the distributions
deviate for any $q$, for example with the marginal energy distribution has 
heavier tails, then the exploration will be slow and the autocorrelations of 
the chain large.}
\label{fig:energy_marginals}
\end{figure}

To quantify the efficacy of the momentum resampling consider $\pi_{E \mid q}$, 
the distribution of energies, $E$, induced by a momentum resampling at 
position $q$.  The closer this distribution is to the marginal energy distribution 
for any $q$, the faster the random walk will explore energies and the smaller 
the autocorrelations we be in the overall Hamiltonian Markov chain
(Figure \ref{fig:energy_marginals}a). Conversely, the more this distribution 
deviates from the marginal energy distribution the less effectively the random 
walk will explore and the larger the autocorrelations will be in the overall chain 
(Figure \ref{fig:energy_marginals}b).

Consequently, the compatibility of the momentum mresampling-induced 
distributions and the marginal energy distribution defines an implicit criterion 
for selecting an optimal cotangent disintegration for a given target distribution.  
There are many ways, however, of quantifying this compatibility in theory and 
in practice, and hence many ways of defining optimality criteria and resulting
diagnostics.

\subsection{General Criteria}

Optimal performance is achieved only when the momentum resampling-induced 
energy distributions are uniformly equal to the marginal energy distribution,
\begin{equation*}
\log \frac{ \dd \pi_{E \mid q} }{ \dd \pi_{E} } = 0, \, \forall q \in Q.
\end{equation*}
Consequently we could quantify the compatibility of the two distributions
with the expectation
\begin{equation*}
\mathbb{E}_{\pi} \! \left[
\log \frac{ \dd \pi_{E \mid q} }{ \dd \pi_{E} } \right],
\end{equation*}
which would vanish only when the cotangent disintegration was optimal.
Because we don't have closed-forms for the densities, however, this
would be infeasible to even estimate in any nontrivial problem.

In practice we want a criterion that is readily estimating using the
Hamiltonian Markov chain itself.  One theoretically appealing choice
is the Bayesian fraction of missing information~\citep{Rubin:2004},
\begin{equation*}
\mathrm{BFMI} = 
\frac{ \mathbb{E}_{\pi} \! 
\left[ \mathrm{Var}_{ \pi_{E \mid q} } \! \left[ E \mid q \right] \right] }
{ \mathrm{Var}_{ \pi_{E} } \! \left[ E \right] },
\end{equation*}
which quantifies how insufficient the energy variation induced by the 
momentum resampling is: in the worst case $\mathrm{BFMI} \rightarrow 0$ 
and the momentum resampling induces very slow exploration across the 
level sets, while in the best case $\mathrm{BFMI} \rightarrow 1$ and the 
momentum resampling effectively generates exact draws from the marginal 
energy distribution.

By construction,
\begin{align*}
\mathrm{Var}_{ \pi_{E \mid q} } \! \left[ E \mid q \right]
&=
\mathrm{Var}_{ \pi_{E \mid q} } \! \left[ \Delta E \mid q \right],
\end{align*}
where $\Delta E$ is the change in energy induced by the momentum
resampling.  Because the momentum resampling does not change
the position, $q$, this can also be interpreted as the change in
kinetic energy, $\Delta E = \Delta K$, which depends only on the
choice of cotangent disintegration, as expected.  Using this latter form
we can then readily estimate the Bayesian fraction of missing information 
using the history of energies in the Hamiltonian Markov chain,
\begin{equation*}
\mathrm{BFMI}
\approx
\widehat{\mathrm{BFMI}} 
\equiv
\frac{ \sum_{n = 1}^{N} \left( E_{n} - E_{n - 1} \right)^{2} }
{ \sum_{n = 0}^{N} \left( E_{n} - \bar{E} \right)^{2} }.
\end{equation*}

In this form the Bayesian fraction of missing information is similar
to the lag-1 autocorrelation of the energies, suggesting that the
effective sample size per transition of the energies, 
$\mathrm{ESS/T} \! \left( E \right)$ might also be a 
useful quantification, with 
$\mathrm{ESS/T} \! \left( E \right) \rightarrow 0$ indicating
a suboptimal cotangent disintegration and 
$\mathrm{ESS/T} \! \left( E \right) \rightarrow 1$ indicating
an optimal one.  This measure also has a strong intuitive appeal -- 
up to the usual regularity conditions the effective sample size 
quantifies the rate of convergence of the marginal random walk
over energies, and hence it directly quantifies the efficacy of the 
exploration induced by the momentum resampling.

Finally, we can also use the the change of energies induced by a
momentum resampling to construct visual criteria.  Averaging the
momentum resampling-induced energy distribution over positions 
gives a marginal distribution over energy variations,
\begin{equation*}
\pi_{\Delta E} \! \left( \Delta E \right) \equiv
\int \pi_{E \mid q} \! \left( \Delta E \mid q \right) \pi \! \left( q \right) \dd q,
\end{equation*}
whose density is readily estimated by histogramming the 
$\left\{ \Delta E_{n} \right\}$ from the Hamiltonian Markov chain.  
We can also estimate the marginal energy density by histogramming 
the $\left\{ E_{n} \right\}$, and then compare the variation of the two
histograms by eye.

The Bayesian fraction of missing information, effective sample size
per transition, and histograms can all be estimated directly from
the history of the Hamiltonian Markov chain, but none of them define 
a criteria that can be explicitly inverted to identify an optimal disintegration.  
Hence in practice they best serve as diagnostics for distinguishing 
suboptimal disintegrations. 

Additionally we must take care when applying these diagnostics as they 
make the strong assumption that the Hamiltonian Markov chain sufficiently 
explores the joint distribution.  In order to improve the robustness of these 
diagnostics, in practice it is best to use them with multiple Markov chains
and monitor additional diagnostics such as 
divergences~\citep{BetancourtEtAl:2014b, BetancourtEtAl:2015} and 
the Gelman-Rubin statistic~\citep{GelmanEtAl:1992}.

\subsection{Gaussian-Euclidean Criteria}

Gaussian-Euclidean cotangent disintegrations are particularly useful as 
they admit some results that can simplify the general criteria introduced
above.

Consider, for example, the random lift onto the cotangent bundle.
For any Gaussian-Euclidean cotangent disintegration, in fact for any 
Gaussian-Riemannian cotangent disintegration, the effective kinetic energy 
introduced by the randomly sampled momentum is distributed according 
to a scaled-$\chi^{2}$ distribution independent of position, 
\begin{equation*}
\widecheck{K} \sim \chi^{2} \! \left( D, 1 / 2 \right),
\end{equation*}
where
\begin{equation*}
\chi^{2} \! \left( x \mid k, \sigma \right)
=
\frac{ \left( 2 \sigma \right)^{-\frac{k}{2}} }{ \Gamma \! \left( \frac{k}{2} \right) }
x^{\frac{k}{2} - 1} e^{-\frac{x}{2 \sigma} }.
\end{equation*}
In general the projection can shed an arbitrarily large amount of effective
kinetic energy, but in equilibrium we'd expect to loose as much energy 
as we gain, hence the change in energies should be distributed as the 
difference between two $\chi^{2} \! \left( D, \frac{1}{2} \right)$ variates
describing the initial and final effective kinetic energies.  As the number of
dimensions, $D$, increases this distribution rapidly converges to a Gaussian 
distribution with zero mean and $D$ variance, so to a good approximation 
we have
\begin{equation*}
\Delta E \sim \mathcal{N} \! \left( 0, D \right),
\end{equation*}
for all positions, $q$. 

In this case the numerator in the Bayesian fraction of missing information
becomes
\begin{equation*}
\mathbb{E}_{\pi} \! 
\left[ \mathrm{Var}_{ \pi_{E \mid q} } \! \left[ E \mid q \right] \right]
=
 \mathbb{E}_{\pi} \! 
\left[ D \right]
= D,
\end{equation*}
and we can quantify the efficacy of the cotangent disintegration simply
by comparing the variance of the marginal energy distribution to the
dimensionality of the target sample space, $D$.

\section{Examples}

In order to demonstrate the utility of these diagnostics, in this section
we'll consider a series of pedagogic examples, starting with identically 
and independently distributed Gaussian and Cauchy distributions and 
then a typical hierarchical model.  

All Markov chains were generated with \textsc{CmdStan}~\citep{CmdStan:2016}, 
using the No-U-Turn sampler~\citep{HoffmanEtAl:2014} to dynamically 
adapt the integration time and a Gaussian-Euclidean cotangent disintegration 
with a diagonal Euclidean metric adapted to the covariance of the target 
distribution.  Unless otherwise specified all other settings were default.  
The exact version of \textsc{CmdStan} can be found at 
\url{https://github.com/stan-dev/cmdstan/commit/ad6177357d4d228e129eefa60c9f399b36e9ac19}, and all Stan programs 
and configurations can be found in the Appendix.

\begin{figure}
\centering
\includegraphics[width=4in]{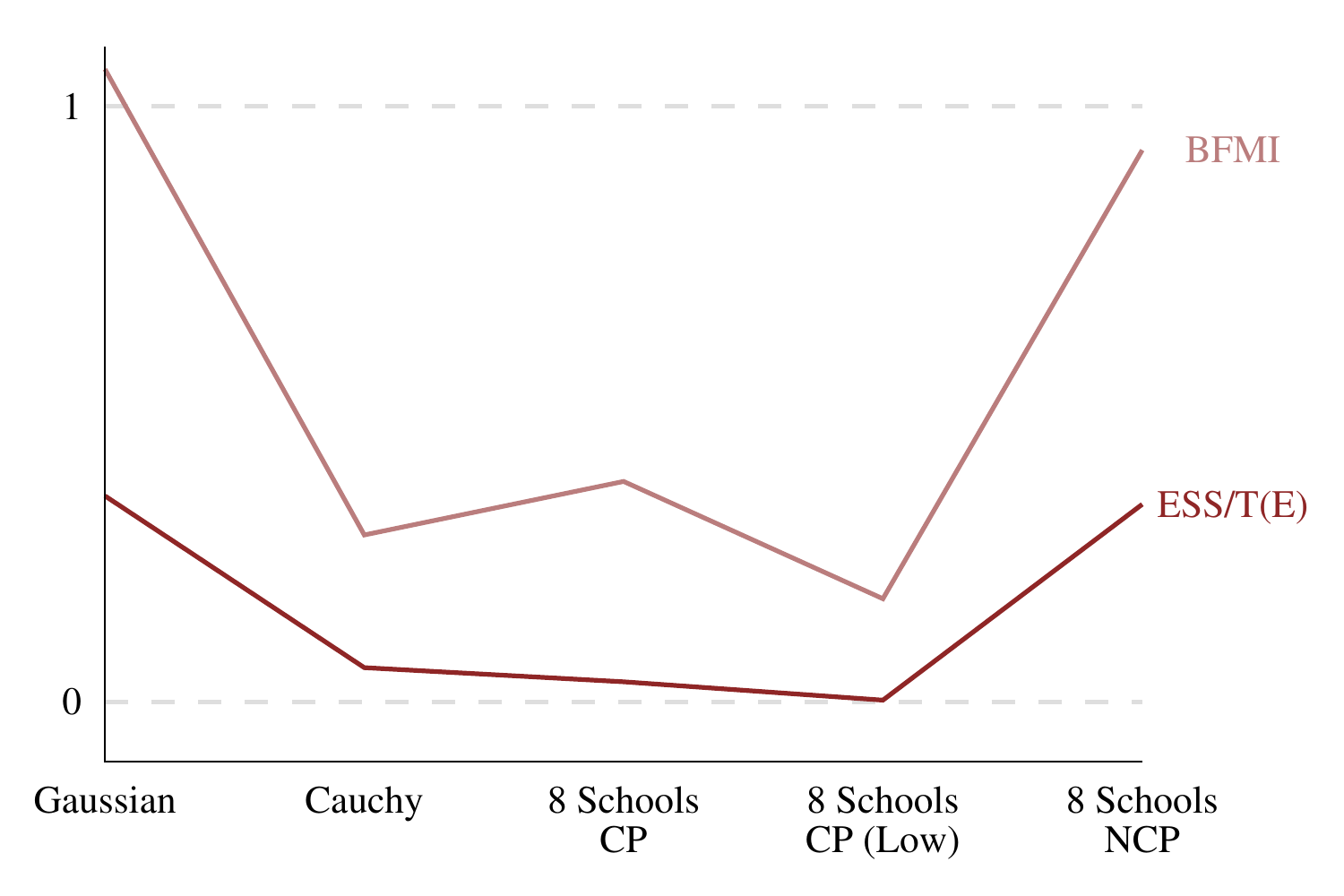}
\caption{Both the Bayesian fraction of mission information, BFMI, 
and effective sample size per transition for the energy ESS/T(E),
quantify the compatibility of a cotangent disintegration with a
given target distribution.  Here a Gaussian-Euclidean cotangent
disintegration works well for both a Gaussian and non-centered eight
schools target, but is less effective for the heavier-tailed Cauchy
and centered eight schools targets.}
\label{fig:num_diagnostics}
\end{figure}

\subsection{Gaussian Target}

\begin{figure}
\centering
\includegraphics[width=4in]{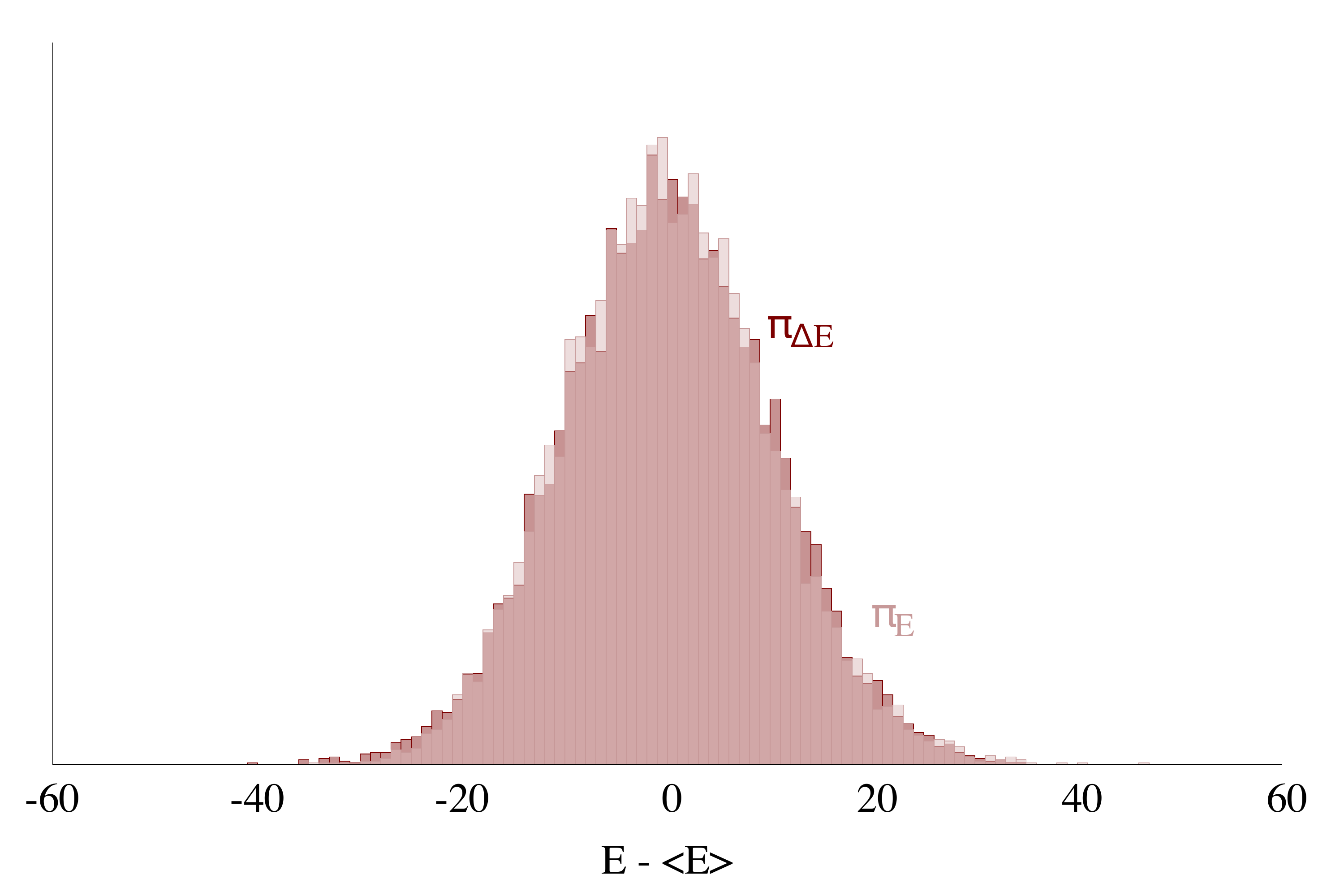}
\caption{A Gaussian-Euclidean cotangent disintegration is well-suited
to a Gaussian target distribution -- at each iteration the momentum
resampling is able to jump the Hamiltonian Markov chain to any relevant
level set.}
\label{fig:gauss_exp}
\end{figure}

Let's first consider a 100-dimensional identically and independently distributed
Gaussian distribution, $q_{n} \sim \mathcal{N} \! \left( 0, 1 \right)$.  Given a
Gaussian-Euclidean cotangent disintegration the marginal energy distribution
reduces to a scaled $\chi^{2}$ distribution, 
\begin{equation*}
E \sim \chi^{2} \! \left( 2 D, \frac{1}{2} \right),
\end{equation*}
which converges to a $\mathcal{N} \! \left( D, D \right)$ with increasing
dimension.  This perfectly matches the expected energy variation, as evident
in both numerical diagnostics (Figure \ref{fig:num_diagnostics})  and visual
diagnostics (Figure \ref{fig:gauss_exp}).

\subsection{Cauchy Target}

\begin{figure}
\centering
\includegraphics[width=4in]{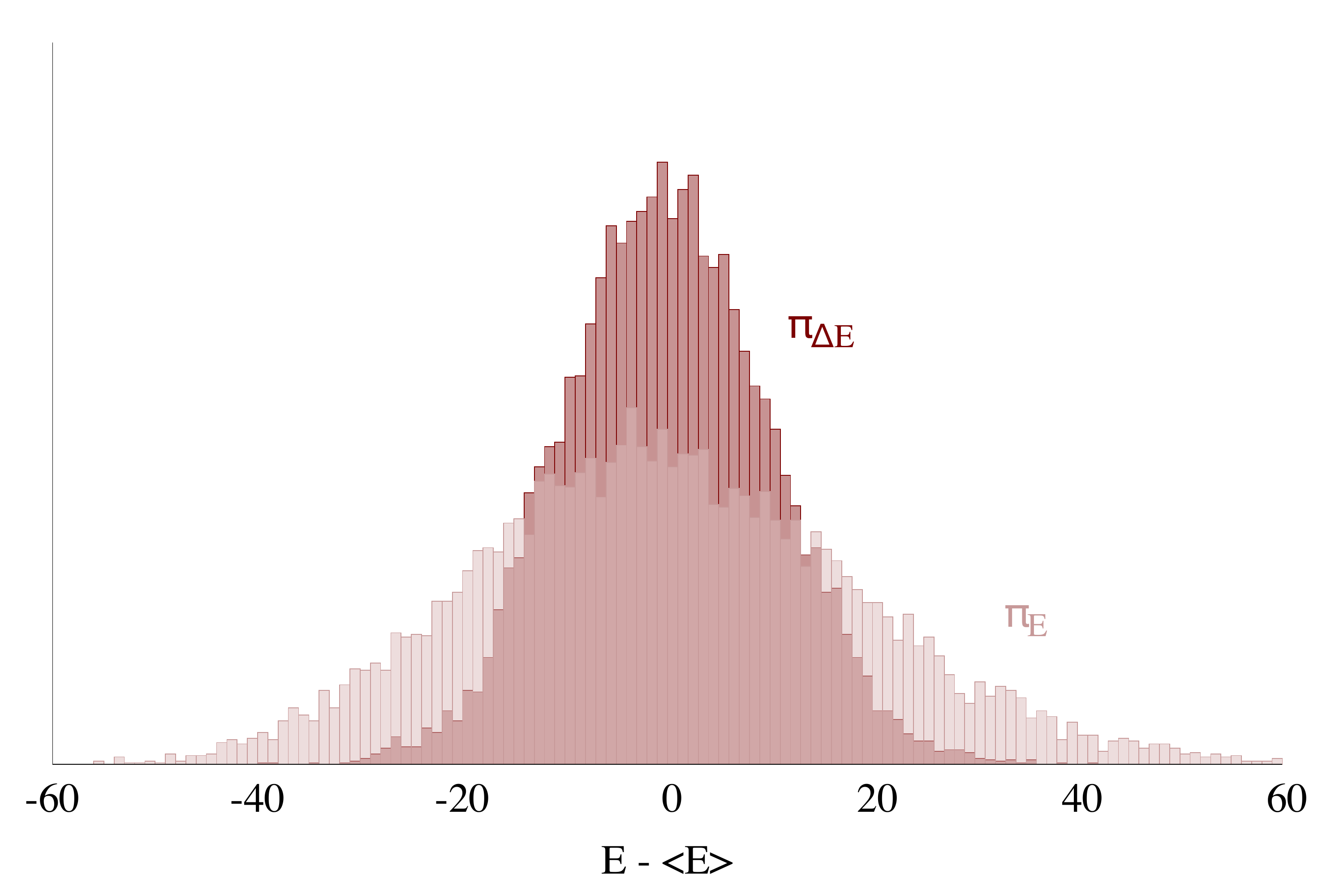}
\caption{The heavy tails of the Cauchy distribution induce a heavy-tailed
marginal energy distribution which limits the efficacy of a Hamiltonian 
Markov chain utilizing the more lightly-tailed energy variation induced by a 
Gaussian-Euclidean cotangent disintegration.}
\label{fig:cauchy_exp}
\end{figure}

For a less compatible pairing consider instead a 100-dimensional identically 
and independently distributed Cauchy distribution, 
$q_{n} \sim \mathcal{C} \! \left( 0, 1 \right)$.  The heavy tails of the Cauchy
distribution induce a marginal energy distribution with heavier tails than
the momentum resampling-induced energy variation.  Consequently each 
transition is limited to only those level sets in close proximity to the initial 
level set, resulting in slower exploration and decreased performance 
(Figures \ref{fig:num_diagnostics}, \ref{fig:cauchy_exp}).  Despite the 
suboptimality of this disintegration, however, the Hamiltonian Markov chain 
is able to explore all relevant energies within only a few transitions and 
ends up performing surprisingly well given the reputation of the Cauchy 
distribution.

\subsection{Hierarchical Target}

Finally, let's consider the eight schools posterior distribution, a relatively 
simple Bayesian hierarchical model that demonstrates both the utility of 
hierarchical modeling as well many of the computational difficulties inherent 
to these models~\citep{Rubin:1981, GelmanEtAl:2014a}.  Here the test taking
performance of eight schools is modeled with individual, centered Gaussian
distributions,
\begin{equation*}
y_{n} \sim \mathcal{N} \! \left( \theta_{n}, \sigma_{n}^{2} \right),
\end{equation*}
where the $\theta_{n}$ are modeled hierarchically,
\begin{align*}
\theta_{n} &\sim \mathcal{N} \! \left( \mu, \tau^{2} \right)
\\
\mu &\sim \mathcal{N} \! \left( 0, 10^{2} \right)
\\
\tau &\sim \text{Half-}\mathcal{C} \! \left( 0, 10 \right),
\end{align*}
and the $\left\{ y_{n}, \sigma_{n} \right\}$ are given as data.

\begin{figure}
\centering
\subfigure[]{ \includegraphics[width=2.5in]{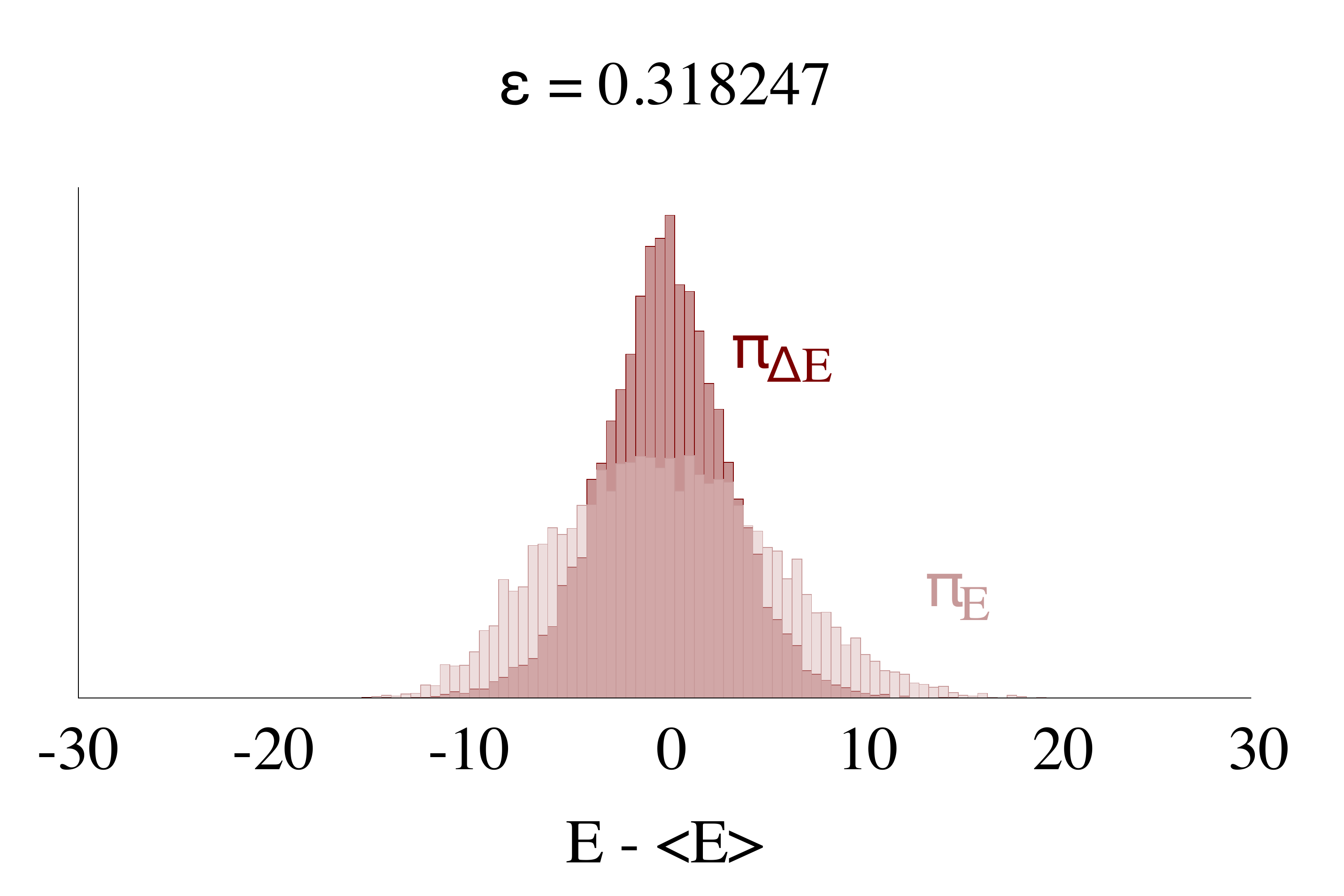} }
\subfigure[]{ \includegraphics[width=2.5in]{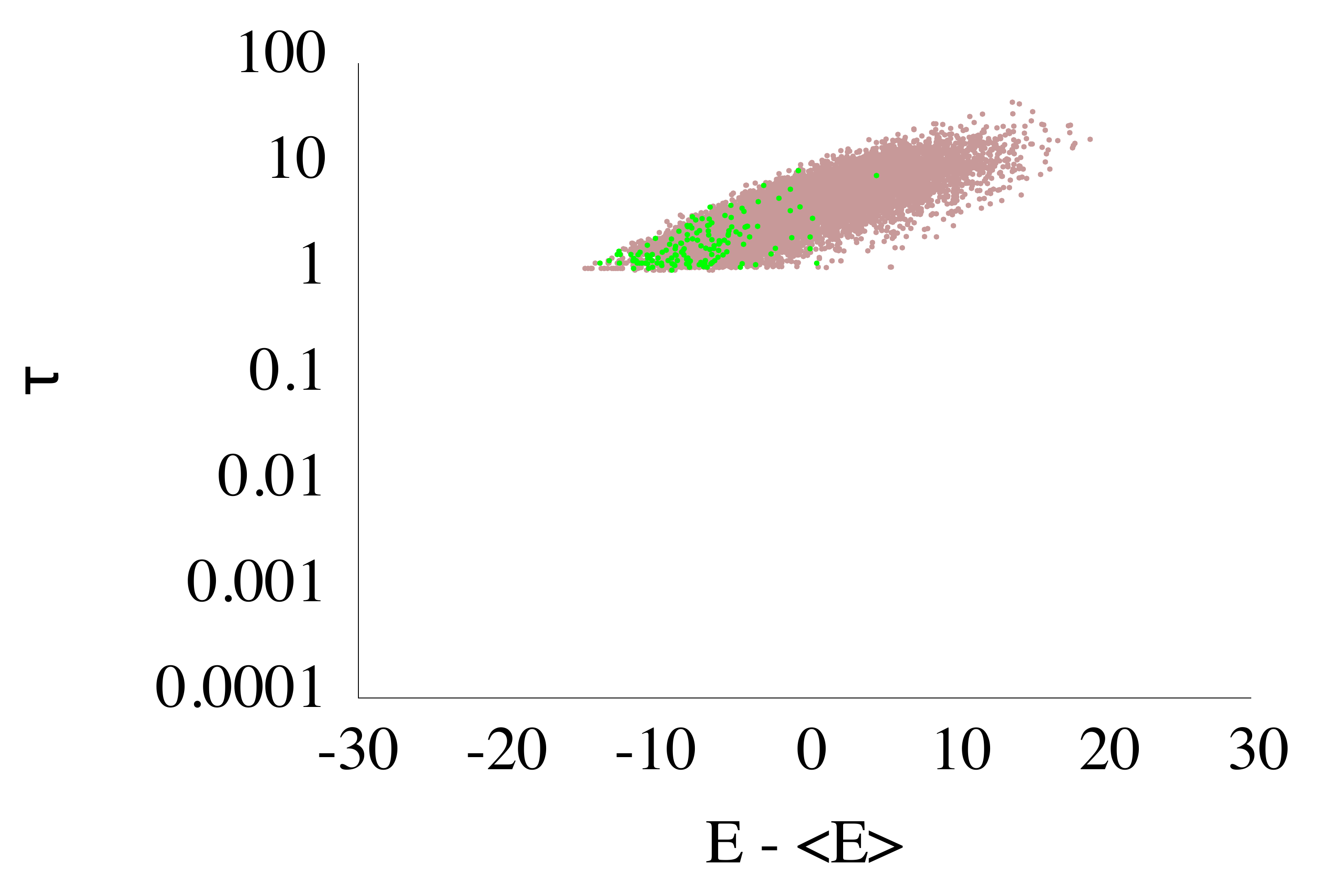} }
\caption{(a) The empirically-derived marginal energy distribution of the
centered eight schools model appears to have mildly-heavy tails,
(b) but this is misleading due to the incomplete exploration of the
tails as indicated by the divergent transitions in the Hamiltonian Markov 
chain shown in green.  With so many divergences, the numerical and visual
diagnostics are suspect.}
\label{fig:energy_schools_cp_exp}
\end{figure}

In the typical centered-parameterization~\citep{PapaspiliopoulosEtAl:2007}
the marginal energy distribution seems to exhibits only mildly-heavy tails 
(Figure \ref{fig:energy_schools_cp_exp}a), but these empirical results are 
misleading.  The problem is that the Hamiltonian Markov chain is not able 
to fully explore the tails of the target distribution, as exhibited by the large 
number of divergences at small $\tau$, 
(Figure \ref{fig:energy_schools_cp_exp}b).

\begin{figure}
\centering
\subfigure[]{ \includegraphics[width=2.5in]{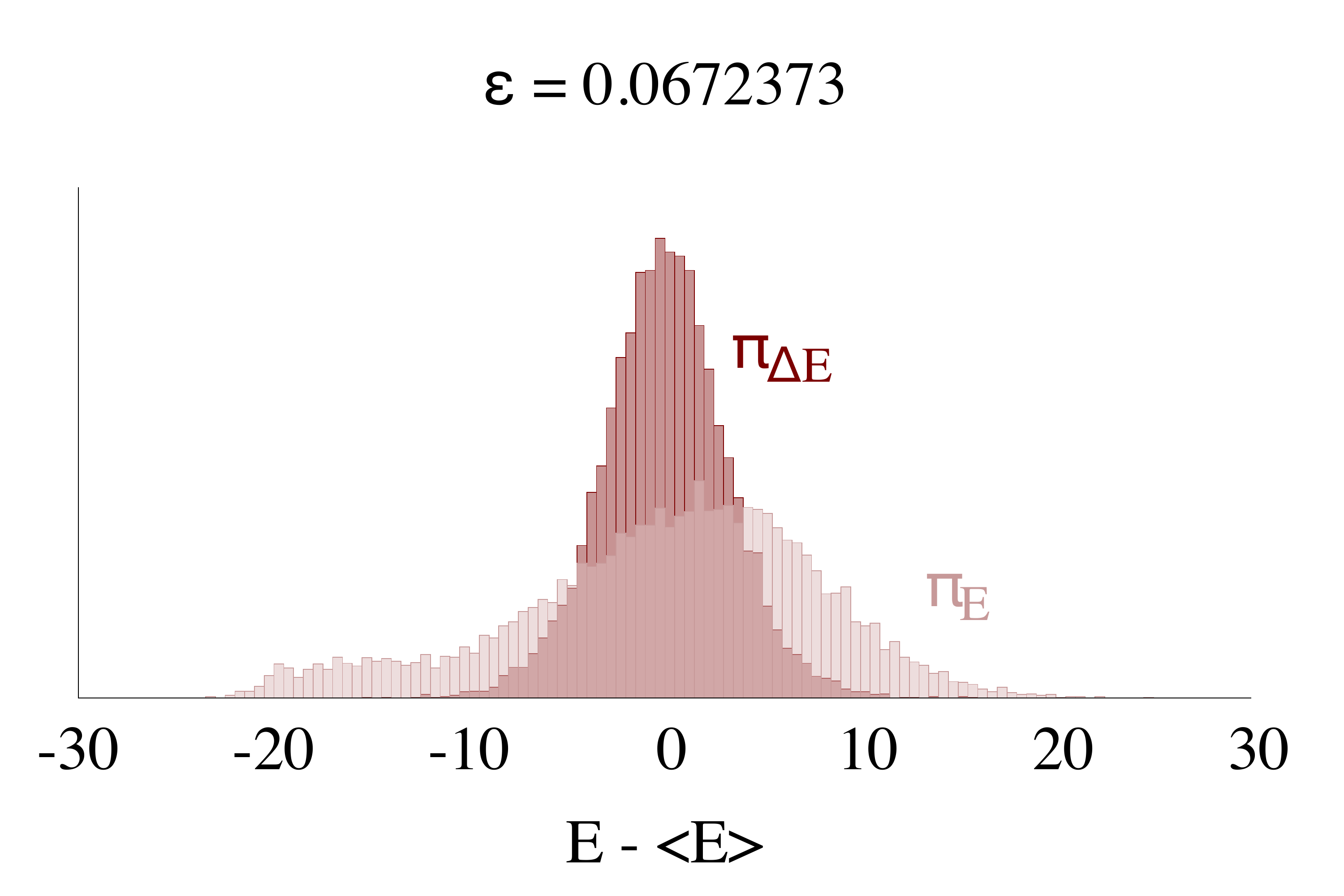} }
\subfigure[]{ \includegraphics[width=2.5in]{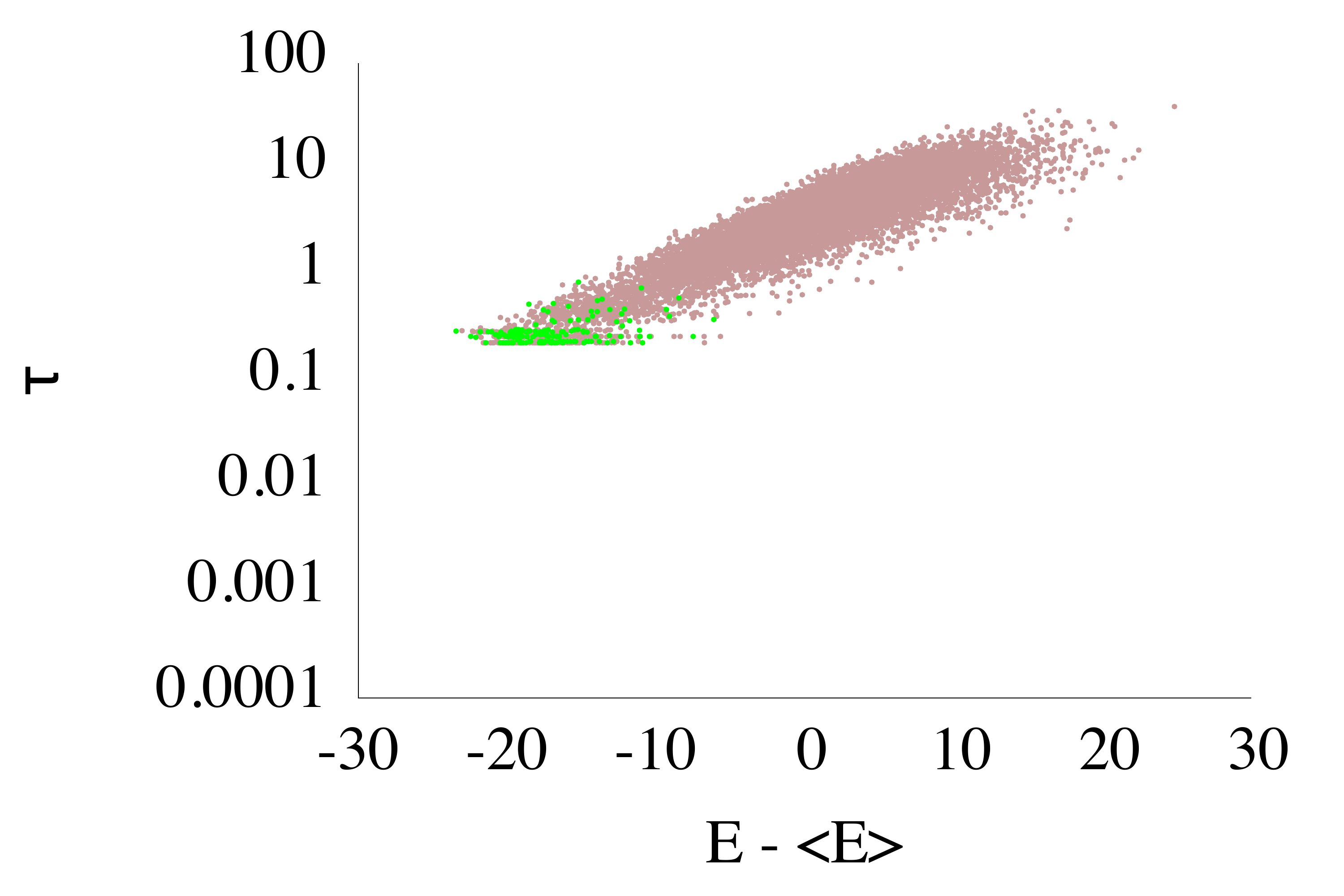} }
\caption{(a) Forcing a smaller step size in the Hamiltonian Markov transition 
admits further exploration of the centered eight schools model and better
exhibits the true heaviness of the marginal energy distribution.  (b) The
persistent divergent transitions (green), however, indicate that the exploration 
is still incomplete.  Regardless of the partial exploration, the 
Gaussian-Euclidean disintegration is suboptimal for this implementation of
the model.}
\label{fig:energy_schools_cp_99_exp}
\end{figure}

Forcing the step size of the numerical integrator to a smaller value
improves the exploration of the tails (Figure \ref{fig:energy_schools_cp_99_exp}b)
and better reveals the true heaviness of the marginal energy distribution
(Figure \ref{fig:energy_schools_cp_99_exp}a), although the exploration
is still incomplete.

\begin{figure}
\centering
\subfigure[]{ \includegraphics[width=2.5in]{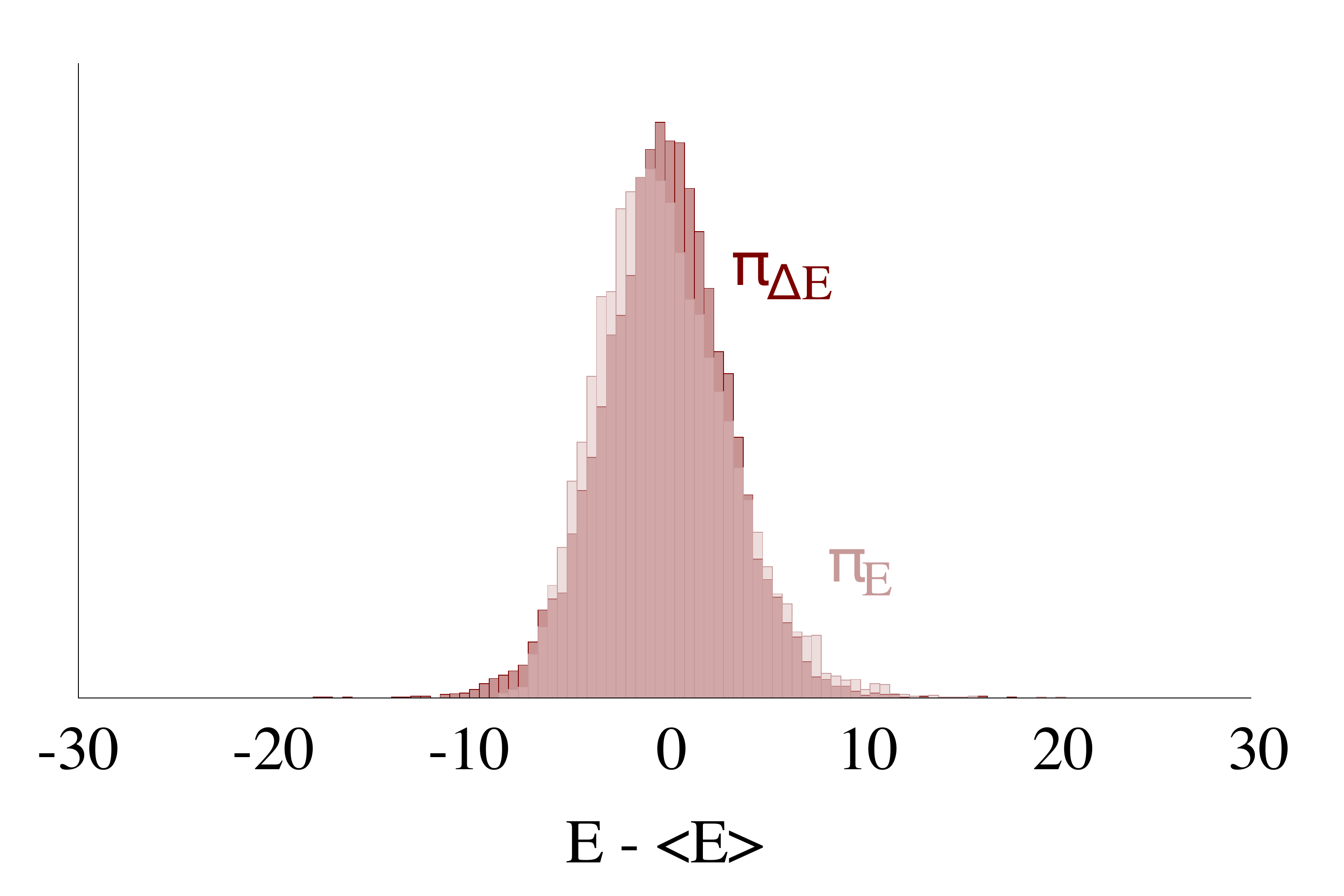} }
\subfigure[]{ \includegraphics[width=2.5in]{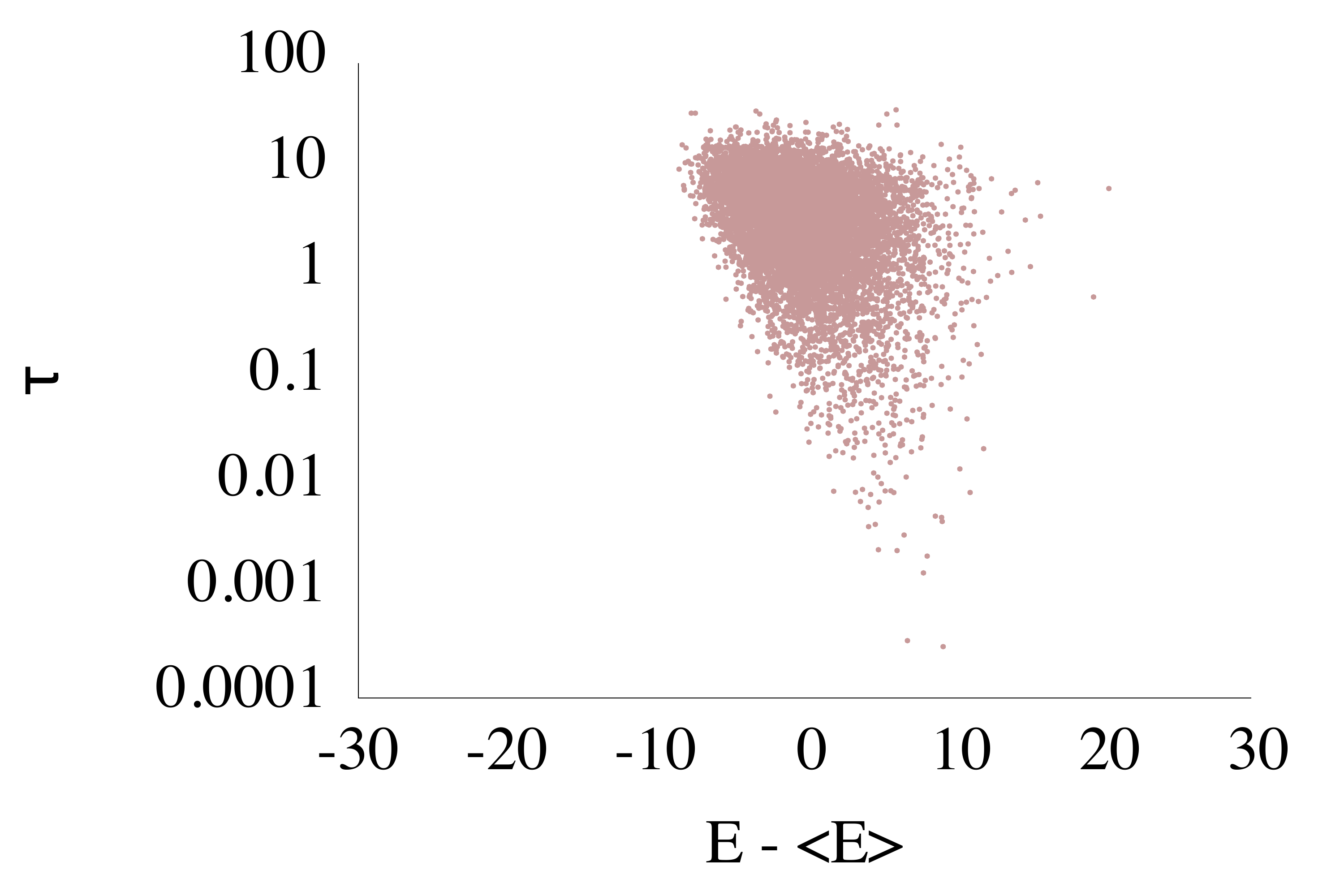} }
\caption{A non-centered implementation of the eight schools model is
not only free of the pathologies that limit the exploration of the centered
implementation but also is a nearly optimal pairing with the 
Gaussian-Euclidean cotangent disintegration.}
\label{fig:energy_schools_ncp_exp}
\end{figure}

In order to completely explore the tails we have to utilize a non-centered
parameterization which explores the hierarchical effects only indirectly,
\begin{align*}
y_{n} &\sim 
\mathcal{N} \! \left( \mu + \tau \cdot \tilde{\theta}_{n}, \sigma_{n}^{2} \right),
\\
\tilde{\theta}_{n} &\sim \mathcal{N} \! \left( 0, 1 \right)
\\
\mu &\sim \mathcal{N} \! \left( 0, 10^{2} \right)
\\
\tau &\sim \text{Half-}\mathcal{C} \! \left( 0, 10 \right).
\end{align*}
Not only does this implementation of the model not suffer from the
heavy tails and pathological curvature of the centered implementation,
the Gaussian-Euclidean cotangent disintegration is a nearly optimal
pairing (Figures \ref{fig:num_diagnostics}, \ref{fig:energy_schools_ncp_exp})

\section{Discussion}

As with any Markov chain Monte Carlo algorithm, the performance of 
Hamiltonian Monte Carlo is limited by its ability to sufficiently explore 
the target distribution.  \cite{LivingstoneEtAl:2016},
for example, demonstrates that both neighborhoods of strong curvature 
and heavy tails limit the exploration of a Hamiltonian Markov chain,
ultimately obstructing geometric ergodicity and the central limit theorems 
needed to ensure robust Markov chain Monte Carlo estimation.

What is unique to Hamiltonian Monte Carlo, however, is its natural
ability to diagnose these pathologies.  Neighborhoods of strong
curvature, for example, can be identified with the divergent transitions
they provoke.  Moreover, heavy tails manifest both in expansive level 
sets and heavy-tailed marginal energy distributions.  When using
dynamic integration time algorithms, the former manifests as long
integration times which are readily reported to users, and we have
seen in this paper that heavy-tailed marginal energy distributions
are straightforward to report both numerically and visually.  These
intrinsic diagnostics make Hamiltonian Monte Carlo extremely
robust, even in the challenging problems at the frontiers of applied statistics.

How to address the pathologies identified by these diagnostics
is another question.  For example, more heavily-tailed cotangent
disintegrations, such as Laplace or even Cauchy disintegrations,
may be useful.  Generalizing from Euclidean to fully Riemannian
disintegrations, for example with the SoftAbs metric~\citep{Betancourt:2013b},
offers another potential strategy.  Within existing tools like \textsc{Stan},
however, perhaps the best way to deal with any identified pathologies is
with alternative implementations, such as the non-centered parameterization
utilized in the eight schools example.

\section{Acknowledgements}

I am grateful to Gareth Roberts for enlightening discussions.  This work
was supported under EPSRC grant EP/J016934/1.

\setcounter{section}{0}
\renewcommand{\thesection}{\Alph{section}}

\section{Stan Programs}

In this section I collect the Stan programs and configurations used in
the examples.

\subsection{Gaussian}

\noindent Configuration:

{\small
\begin{verbatim}
./gauss sample num_samples=10000 random seed=2983157687
\end{verbatim}
}

\noindent Stan Program:

{\small
\begin{verbatim}
parameters {
  real x[100];
}
model {
  x ~ normal(0, 1);
}
\end{verbatim}
}

\subsection{Cauchy}

\noindent Configuration:

{\small
\begin{verbatim}
./cauchy sample num_samples=10000 random seed=2983158736
\end{verbatim}
}

\noindent Stan Program:

{\small
\begin{verbatim}
parameters {
  real x[100];
}
model {
  x ~ cauchy(0, 1);
}

\end{verbatim}
}

\subsection{Centered Eight Schools}

\noindent Nominal Configuration:

{\small
\begin{verbatim}
./eight_schools_cp sample num_samples=10000
data file=eight_schools.data.R random seed=483892929
\end{verbatim}
}

\noindent Small Stepsize Configuration:

{\small
\begin{verbatim}
./eight_schools_cp sample num_samples=10000 adapt delta=0.99
data file=eight_schools.data.R random seed=483892929
\end{verbatim}
}

\noindent Stan Program:

{\small
\begin{verbatim}
data {
  int<lower=0> J;
  real y[J];
  real<lower=0> sigma[J];
}
parameters {
  real mu;
  real<lower=0> tau;
  real theta[J];
}
model {
  mu ~ normal(0, 10);
  tau ~ cauchy(0, 10);
  theta ~ normal(mu, tau);
  y ~ normal(theta, sigma);
}
\end{verbatim}
}

\subsection{Noncentered Eight Schools}

\noindent Configuration:

{\small
\begin{verbatim}
./eight_schools_ncp sample num_samples=10000
data file=eight_schools.data.R random seed=483892929
\end{verbatim}
}

\noindent Stan Program:

{\small
\begin{verbatim}
data {
  int<lower=0> J;
  real y[J];
  real<lower=0> sigma[J];
}
parameters {
  real mu;
  real<lower=0> tau;
  real theta_tilde[J];
}
transformed parameters {
  real theta[J];
  for (j in 1:J)
    theta[j] = mu + tau * theta_tilde[j];
}
model {
  mu ~ normal(0, 10);
  tau ~ cauchy(0, 10);
  theta_tilde ~ normal(0, 1);
  y ~ normal(theta, sigma);
}
\end{verbatim}
}

\bibliography{energy_diagnostic}
\bibliographystyle{imsart-nameyear}

\end{document}